\documentstyle[twocolumn,seceq,epsf]{jpsj}

\def\runtitle{}
\def\runauthor{Satoshi {\sc Sawai} and Yoji {\sc Aizawa}}

\title
{
Coupled Oscillators with Chemotaxis
}

\author
{ 
Satoshi {\sc Sawai}\footnote{Present address: Research Institute of
Electrical Communication, Tohoku University, Sendai 980-8577.}
and Yoji {\sc Aizawa}
}

\inst
{
Department of Applied Physics, Waseda University, Tokyo 169-8555
}

\recdate
{
May 18, 1998
}

\abst
{
A simple coupled oscillator system with chemotaxis is introduced
to study morphogenesis of cellular slime molds.  The model
successfuly explains the migration of pseudoplasmodium which has
been experimentally predicted to be lead by cells with higher
intrinsic frequencies.  Results obtained predict that its velocity
attains its maximum value in the interface region between total
locking and partial locking and also suggest possible roles
played by partial synchrony during multicellular development.   
}

\kword
{
coupled oscillators, cellular slime mold, morphogenesis
}

\begin{document}
\sloppy
\maketitle

\section{Introduction}

Inspired by the collective behavior and rhythmicity of biological
systems, synchronization of coupled limit-cycle oscillators with
a frequency distribution has been studied using a system with
all-to-all interaction in the following form~\cite{rf:1,rf:2,rf:3}, 
\begin{equation}
\dot{W}_j=({\rm i}\omega_j+1-|W_j|^2)W_j+\frac{\epsilon}{N}\sum_{k=1}^N
(W_k-W_j), \label{eq:1}
\end{equation}
where $\omega_j$ is an intrinsic frequency of the oscillator $j$,
$W_j$ is a complex variable and $(\dot{})=\frac{d}{dt}()$.  The
system has served well for theoretical studies on frequency
entrainment and critical behaviors in a system far from
equilibrium.  Despite its original aim, however, very little has
been discussed on the system's applicability and a connection to
biological systems.     

Development of the cellular slime mold {\it Dictyostelium discoideum}
provides us with an ideal example that needs to be investigated
in this respect, since it's chemoattractant secretion exhibits  
limit-cycle oscillations in the vicinity of a Hopf bifurcation
point~\cite{rf:4} with a difference in intrinsic frequencies and
synchronization~\cite{rf:5}.  

These eukaryotic cells feed on bacteria and grow by binary
fission.  Deprived of food, they initiate aggregation by emitting
cAMP as a chemoattractant while simultaneously making a directed
motion against its surrounding gradient.  Each aggregation
territory consists of $10^3$ to $10^5$ amoebae that together form
a spherical mound whose nipple like apex zone consists of
differentiated prestalk cells.  The tipped mound elongates
vertically to form a slug like pseudoplasmodium which migrates to
a suitable environment where it forms a fruiting
body~\cite{rf:6,rf:7}.     

To study a self-organizing process of cells in such systems,
one must understand the nature of collective behavior in a
population of motile oscillators.  In the following, we first
introduce a system that incorporates chemotaxis into
eq. (\ref{eq:1}) and present the system's overall behavior.  Then
we discuss our results and their implications on development of a
cellular slime mold.    

\section{Equations}

The system is derived from a linear diffusion equation for two
chemical species denoted by a complex variable $Z({\mib r},t)$
and equations for chemotaxis of cell $j$ at ${\mib r}={\mib
r}_j$.  They are 

\begin{eqnarray}
\partial_t Z({\mib r},t) &=& D\nabla^2Z({\mib r},t), \label{eq:2.1} \\
\dot{\mib r}_j(t) &=& \tilde{\alpha} \nabla{\rm Re}
Z({\mib r},t),   \label{eq:2.2}
\end{eqnarray} 
where $D$ is a diffusion constant, $\tilde{\alpha}$ is a
chemotactic sensitivity coefficient.

Cell $j~(=1,2,\cdot \cdot \cdot,N)$ is represented by a region [$0 \leq |{\mib
r}-{\mib r}_j| \leq r_0$] on which a boundary condition
representing the metabolism ${\cal F}$ of intracellular chemical
species in a complex variable $W_j(t)$ is imposed.  The boundary
conditions are expressed as
\begin{eqnarray}
\lim_{r \rightarrow \infty} Z({\mib r},t) &=& 0, \\
Z({\mib r}_j+{\mib r}_0,t) &=&
W_j(t)+C, \label{eq:boundary}\\ 
\dot{W}_j(t)&=& {\cal F}(W_j)+\psi(\{W_k\}). \label{eq:2.5}
\end{eqnarray}
Here $\psi(\{W_k\})$ is an abbreviation for the interaction term
$\psi(\{W_1, W_2, \cdot \cdot \cdot, W_N\})$, and $C$ is a constant background level of $Z({\mib r},t)$.  
Solving eq. (\ref{eq:2.1}) under $N$ boundaries
(\ref{eq:boundary}) would yield a field $Z({\mib r},t)$ in an
integral equation 
\begin{equation}
Z({\mib r},t) = \sum_{k=1}^{N}
\int_{0}^t\Phi(|{\mib r}(t)-{\mib
r}_k(\tau)|,t-\tau)(W_k(\tau)+C) {\rm d}\tau,
\end{equation} 
where the diffusion kernel $\Phi$ is either a Gaussian or
Bessel-type function according to the dimension of ${\mib r}$.  

In order to simplify the system, let us assume $\Phi(|{\mib r}(t)-{\mib r}_k(\tau)|,t-\tau) \rightarrow
\delta(t-\tau)$ in the limit of $|{\mib r}-{\mib r}_k|\rightarrow
0$.  Assuming we had ${\cal F}$ that yields the Hopf bifurcation
normal form under a mean-field coupling $\psi(\{W_k\})\propto (Z({\mib
r}_j,t)-Z({\mib r}_j+{\mib r}_0,t))$; eqs.(\ref{eq:2.2}) and
(\ref{eq:2.5}) could be rewritten as     

\begin{full}
\begin{eqnarray}
\dot{W}_j(t) &=& {\rm i}\omega_j W_j(t)+(\lambda-|W_j(t)|^2)W_j(t) 
+\frac{\epsilon}{N-1}\sum_{k \neq j}^{N}
\int_{0}^t\Phi(|{\mib r}_j(t)-{\mib
r}_k(\tau)|,t-\tau)W_k(\tau) {\rm d}\tau, \label{eq:2.7} \\
\dot{\mib r}_j &=& \frac{\alpha}{N-1}\sum_{k \neq j}^{N}\int_{0}^t \nabla
\Phi(|{\mib r}_j(t)-{\mib r}_k(\tau)|,t-\tau) {\rm
Re}(W_k(\tau)+C) {\rm d}\tau, 
\label{eq:2.8} 
\end{eqnarray}
\noindent where $\omega_j$ and $\lambda$ independently determine the
intrinsic frequency and the amplitude of the oscillation.  In a
weakly coupled regime, we could neglect the amplitude effect,
therefore allowing  eqs.(\ref{eq:2.7}) and (\ref{eq:2.8}) to be
further reduced to a coupled phase model in the following form,    
\begin{eqnarray}
\dot{\phi_j}(t) &=& \omega_j+\frac{\epsilon}{N-1}\sum_{k \neq
j}^{N}\int_{0}^t\Phi(|{\mib r}_j(t)-{\mib r}_k(\tau)|,t-\tau)
\sin(\phi_k(\tau)-\phi_j(t)) {\rm d}\tau, 
\label{eq:2.9} \\
\dot{\mib r}_j &=&
\frac{\gamma(\phi_j)}{N-1}\sum_{k \neq
j}^{N}\int_{0}^t\nabla\Phi(|{\mib r}_j(t)-{\mib
r}_k(\tau)|,t-\tau)
(\cos\phi_k(\tau)+c) {\rm d}\tau+{\mib v}_d, \label{eq:2.10} 
\end{eqnarray}
\end{full}
\noindent where $c$ is the real part of $C$.  In the transformation from
eq. (\ref{eq:2.7}) to eq. (\ref{eq:2.9}), we have fixed
$\lambda=1$.  In addition, the deceleration term
and a periodic change of chemotaxis sensitivity were added to our 
previous model~\cite{rf:8}.  Here, the sensitivity function $\gamma$
and the short range interaction $v_d$ are defined by    
\begin{eqnarray}
\gamma(\phi)&=&\alpha(1-\kappa \cos \phi), \\
{\mib v}_d &=&-\sum_{n\neq j}^N
\frac{\beta}{(|{\mib r}_j(t)-{\mib r}_n(t)|-2r_0)^2} \nonumber \\
& \cdot & \frac{{\mib r}_j(t)-{\mib r}_n(t)}{|{\mib r}_j(t)-{\mib r}_n(t)|}, 
\end{eqnarray}
where $\beta$ and $\kappa$ are both positive constants.

By imposing a boundary condition on each cell that 
is in the form of an ordinary differential equation,
the spatially extended system in eqs.~(\ref{eq:2.1}) and (\ref{eq:2.2}) is
now reduced to a set of integro-differential equations that
tracks the time development of those cell boundaries.  Notice that
eq. (\ref{eq:2.9}) provides us with a general scheme that incorporates
spatial dependency in the well-studied phase
model~\cite{rf:9,rf:10}.  This is a novel approach to
reaction-diffusion systems applicable to those that exhibit
nonlinearity localized on boundaries.     

\section{Method}

For the sake of numerical analysis, the kernel is simplified down to a
point where it is not distinguishable from an one dimensional kernel
except by the factor of $\sqrt{\frac{r_0}{r}}$ and a shift in the
origin of the exponent.  Furthermore, it is multiplied by a step function
which roughly incorporates the effect of degradation of the
chemoattractant by the enzyme in the extracellular substratum.
Therefore $\Delta t$ could be considered as an average life span 
of the chemoattractant.  The precise form used in the
calculations is as follows: 
\begin{equation}
\Phi(r,t-\tau)=\frac{\Theta(\tau-t+\Delta t)(r-r_0)}{\sqrt{4\pi
D(t-\tau)^3}}{\sqrt{\frac{r_0}{r}}} {\rm e}^{-\frac{r^2}{4D(t-\tau)}},
\end{equation}
where $\Theta(t)$ denotes a Heaviside step function.

Numerical studies on eqs. (\ref{eq:2.9}) and (\ref{eq:2.10}) were
performed using the fourth-order Runge-Kutta method in addition
to the semiopen formula~\cite{Abramowitz} for the integral terms.
Parameters $D$, $r_0$ and $\kappa$ were fixed at unity throughout latter
calculations.  Other parameters are $c=2.0$,
$\omega_j=1.0+(j-1)\Delta \omega$, $\beta=0.01$, $\Delta t=2.0$
unless stated otherwise.      

We have employed constant initial values for the interval of
$[-\Delta t,0]$, and all results were obtained from a fixed step
size of h=0.01.  Some calculations where $N=2$ were checked for
accuracy using h=0.001 and h=0.0001.    

\section{Migration in $N=2$}
We will first describe the simplest case where $N=2$ to
characterize attractors in the system.  In order to carry out a
steady-state approximation, chemotaxis will be confined to one
dimension, and the adaptation will also not be considered
($\kappa=0$).     

Suppose oscillators were entrained with a constant phase
difference $\psi=\phi_1-\phi_2$.  When $x_2-x_1=2r_0=const.$, the
equation describing the position of a centroid $x_c=(x_1+x_2)/2$ could
be approximated by
\begin{equation}
\dot{x}_c \simeq \psi \alpha \int_0^t 
\nabla \Phi(2r_0,\tau') \sin\phi_1(t-\tau'){\rm d}\tau',
\label{eq:velpsi}
\end{equation}
where we have neglected the second- and higher-order terms of $\psi$ and had a
change in variable from $\tau$ to $\tau'=t-\tau$.  The mean cluster velocity
$v_c=<\dot{x_c}>$ therefore would be proportional to $\psi$.  

We see that $v_c$ is also proportional to $\Delta \omega$
from the locking condition $\dot{\psi}=0$.  Applying
the same approximation as in eq. (\ref{eq:velpsi}), 
\begin{equation}
\dot{\psi}\simeq-\Delta\omega-{2\psi\epsilon}\int_0^t\Phi(2r_0,\tau')\cos\omega^*\tau'{\rm d}\tau', \label{eq:clustervel}
\end{equation}
where $\omega^*$ is the entrained frequency.  When the phase is locked,
\begin{equation}
\psi\simeq-\frac{\Delta \omega }{2\epsilon
\int_0^t\Phi(2r_0,\tau')\cos\omega^*\tau'{\rm d}\tau'} \label{eq:locking}
\end{equation}
must be satisfied.  From eqs.(\ref{eq:velpsi}) and
(\ref{eq:locking}), one obtains $|v_c|\propto \alpha \Delta
\omega/\epsilon$. 

\begin{figure}[t]
\begin{center}
\leavevmode
\epsfxsize=6.5cm
\epsfbox{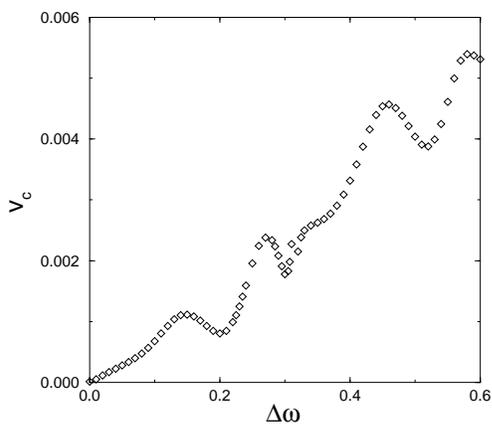}
\end{center}
\caption{$\Delta \omega$ and the cluster velocity $v_c$ ($\omega_1=1.0,
\omega_2=\omega_1+\Delta \omega, \epsilon=1.0$ and $\alpha=1.0$).
$v_c$ was obtained using the method of least squares.}   
\label{fig:vel99020731}
\end{figure}

\begin{figure}[t]
\begin{center}
\leavevmode
\epsfxsize=6.5cm
\epsfbox{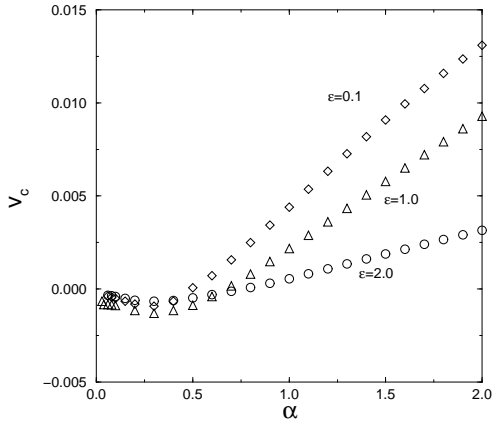}
\end{center}
\caption{$\alpha$ and the cluster velocity for
different $\epsilon$~($\omega_1=1.0, \omega_2=1.1$ and
$\alpha=1.0$).  The initial condition was chosen so that $x_1(0)<x_2(0)$. }  
\label{fig:vel99030611}
\end{figure}

The parameter dependencies given here were confirmed by numerical
simulations for $N=2$ with $\kappa=1.0$.  For a sufficiently large
$\alpha$, the oscillator with larger $\omega_j$ is advanced in
phase and leads migration as one can see in
Figs.~\ref{fig:vel99020731} and \ref{fig:vel99030611}. Not
only $\alpha$ but $\Delta \omega$ also increases the cluster
velocity $v_c$.  Note that there is no net migration of a cluster
when $\Delta \omega=0$.  Figure \ref{fig:vel99030611} also
indicates  that there is velocity proportional to $\alpha$ (but
not to 1/$\epsilon$) even when frequency is not entrained
($\epsilon=0.1$) due to the deceleration term which was not
present in Aizawa-Kohyama (A-K) model discussed
previously~\cite{rf:8}.            

\section{Numerical Results for $N=20$}

\begin{figure}[t]
\begin{center}
\leavevmode
\epsfxsize=6cm
\epsfbox{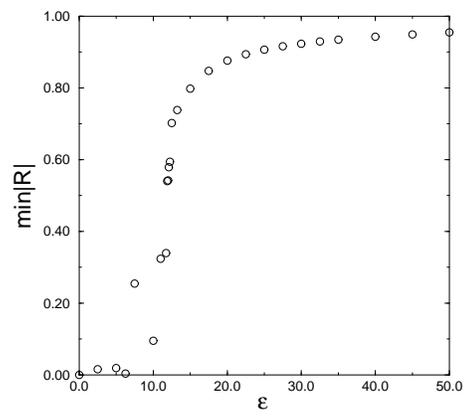}
\end{center}
\caption{$\epsilon$ and the order parameter $R$~($\alpha=1.0$ and $\Delta \omega=0.01$).  The minimum is plotted to withdraw fluctuations due to small $N$.}   
\label{fig:ordereps2}
\end{figure}

In order to characterize coherence in the phase dynamics, an
order parameter $R=\frac{1}{N}\sum_{j=1}^N e^{i\phi_j}$ is plotted against $\epsilon$ in 
Fig.~\ref{fig:ordereps2}.  Since we are dealing with a very small
number of $N$, $|R|$ does not approach zero as $\epsilon \to 0$.
Note that in Fig.~\ref{fig:ordereps2}, the minimum of $|R|$ was
plotted instead of, for example, $<|R|>$~($<>$ denotes time averaging).  

\begin{figure}[t]
\begin{center}
\leavevmode
\epsfxsize=6cm
\epsfbox{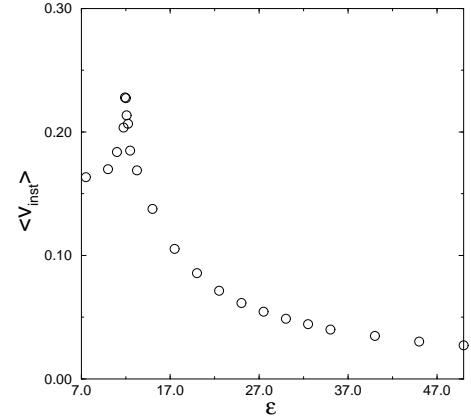}
\end{center}
\caption{$\epsilon$ and cluster velocity~($\alpha=1.0$ and
$\Delta \omega=0.01$).}    
\label{ordereps2-1-3}
\end{figure}

A cluster shows a directed migration when oscillators are
entrained into a common frequency.  The instantaneous velocity of a
centroid defined by ${\mib r}_c=\frac{1}{N}\sum_{j=1}^N
{\mib r}_j$ is plotted against $\epsilon$ in
Fig.~\ref{ordereps2-1-3}.  The apparent discontinuity at the
onset of total entrainment suggests that the cluster velocity, too,
may be taken as an order parameter.   

The relation between the polarity of a cluster and the migration
direction could be understood by introducing $Z(\theta)$ which is 
defined by  
\begin{equation}
Z(\theta)=\sum_{k=1}^{N}km_k(\theta),
\end{equation}
where $m_k$s are natural numbers $\{1,2,3, \cdot \cdot \cdot,
N\}$ representing the distance of oscillator $j$ from a given
point outside the cluster $(x,y)=(r\cos\theta,r\sin\theta)$; $1$
is assigned to the closest and $20$ to the furthest oscillator.
From the above definition, $Z(\theta)$ takes the maximum value in
the direction $\theta$ (measured from the centroid) where
oscillators with a smaller $\omega_j$ are positioned.    

\begin{figure}[t]
\begin{center}
\leavevmode
\epsfxsize=6cm
\epsfbox{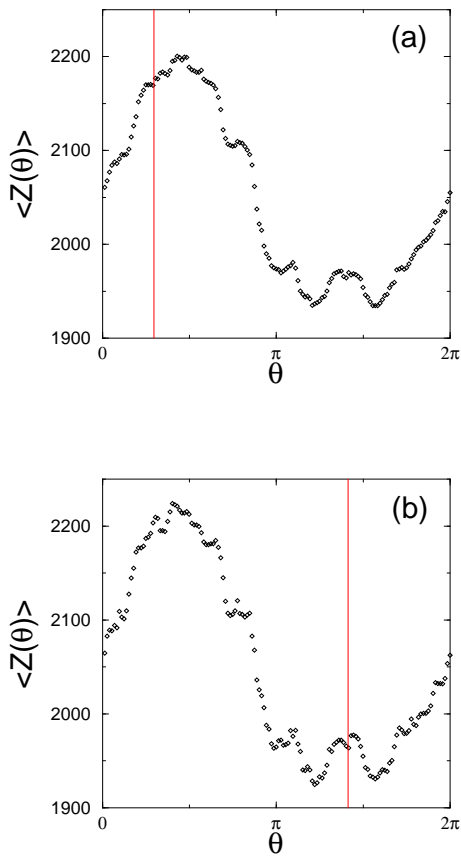}
\end{center}
\caption{Cluster polarity and migration direction
($\epsilon=30.0$, $\alpha=1.0, \Delta \omega=0.01$).  Vertical
lines indicate the direction of cluster migration.  Other
parameters are $\Delta t=2.0$~(a) and $\Delta t=6.0$~(b). } 
\label{fig:zetaeps}
\end{figure}

Figure \ref{fig:zetaeps} displays $Z(\theta)$ and the direction of
cluster migration.  When $\epsilon$ and $\Delta t$ are small but
large enough to synchronize oscillators, a cluster heads
toward the direction where more oscillators with smaller $\omega_j$
are located.  The orientation reverses as both $\epsilon$
and $\Delta t$ are increased so as to make the coupling long
range.  

\section{Discussion}

A simple coupled oscillator model of cell aggregates was derived
from a linear diffusion equation with time-dependent boundaries.
The approximation in $N=2$ and numerical analysis carried out for
$N=20$ revealed that synchrony in such a population of
oscillators with chemotaxis results in migration as a whole.  In
addition to the migration direction that agrees with an
experimental prediction, there are some implications from our
present work concerning the possible order parameter.   

We showed in Fig.~\ref{fig:zetaeps} that in the case of $N=20$, oscillators with larger $\omega_j$ also lead the
cluster translocation if $\Delta t$ is sufficiently large.  In
the light of the prediction from suspension
experiments~\cite{rf:5} that cells with higher frequency
constitute the anterior of a slug, it may be concluded that cell to cell
interaction is not local but rather long range.  The
opposite migratory direction predicted by a small $\Delta t$
suggests that a reverse flow such as the one exhibited by
subpopulation of cells at the onset of
culmination~\cite{rf:11} could be realized without any
secondary chemoattractant.  It would be 
interesting to see whether cells make use of such effect caused
by different coupling ranges.  This could be controlled either 
by the extracellular enzyme concentration or by the surface receptor
occupancy.      

The relation between the coupling strength $\epsilon$ and the
instantaneous cluster velocity has partly confirmed the result
obtained by A-K model~\cite{rf:8} with its peak centered
around the critical region.  Below $\epsilon=7.5$, $v_c$ continues to
fluctuate with its direction not fixed which prevents us
from obtaining its mean.

We have observed that in a partially synchronized state,
locked oscillators emerge specifically in the anterior section
of a cluster (data not shown) which implies that
desynchronization plays a possibile role in cell-type
differentiation.  In general, $f^{-\mu}$ fluctuations are to be
observed in such interface region.  A detailed analysis on such
fluctuation and its effect will be provided elsewhere.  

Due to the coupling and chemotaxis in the form of integral
equation, we have confined our present report on results obtained
in the case of $N=20$ with fixed $r_0$ and $D$.  The diffusion characteristic makes it difficult to vary the ratio
$r_0^2/D$ since decreasing it requires smaller step size $h$
and increasing it requires larger $\Delta t$.  Future work will
also address the construction of a simpler scheme that
makes the analysis more feasible.   

\section*{Acknowledgements}
We thank Prof.~Y.~Sawada for valuable discussions and remarks.

\end{document}